\documentclass[12pt]{article}
\usepackage{amsmath,epsfig,amssymb,bbm}

\newcommand{\jtheta}[1]{\vartheta \begin{bmatrix} #1 \end{bmatrix}}

\newcommand{\Z}{{\bf Z}}
\newcommand{\1}{{\mathbbm{1}}}

\renewcommand{\Im}{{\rm Im}}

\begin{document}

\title{\vbox{
\baselineskip 14pt
\hfill \hbox{\normalsize WU-HEP-09-01} \\
\hfill \hbox{\normalsize KUNS-2202}\\
\hfill \hbox{\normalsize YITP-09-29} } \vskip 2cm
\bf Non-Abelian Discrete Flavor Symmetries from Magnetized/Intersecting 
Brane Models \vskip 0.5cm
}
\author{Hiroyuki~Abe$^{1,}$\footnote{email:
 abe@waseda.jp}, \
Kang-Sin~Choi$^{2,}$\footnote{email:
  kschoi@gauge.scphys.kyoto-u.ac.jp}, \
Tatsuo~Kobayashi$^{2,}$\footnote{
email: kobayash@gauge.scphys.kyoto-u.ac.jp} \ \\ and \
Hiroshi~Ohki$^{2,3,}$\footnote{email: ohki@scphys.kyoto-u.ac.jp
}\\*[20pt]
$^1${\it \normalsize
Department of Physics, Waseda University, Tokyo 169-8555, Japan} \\
$^2${\it \normalsize
Department of Physics, Kyoto University,
Kyoto 606-8502, Japan} \\
$^3${\it \normalsize 
Yukawa Institute for Theoretical Physics, Kyoto University, 
Kyoto 606-8502, Japan}
}

\date{}

\maketitle
\thispagestyle{empty}

\begin{abstract}
We study non-abelian discrete flavor symmetries, 
which can appear in magnetized brane models.
For example, $D_4$, $\Delta(27)$ and $\Delta(54)$ 
can appear and matter fields with 
several representations can appear.
We also study the orbifold background, 
where non-abelian flavor symmetries are broken 
in a certain way.
\end{abstract}

\newpage

\section{Introduction}

It is one of important issues to study 
the origin of the quark/lepton flavor structure; 
why there are three generations, 
why the hierarchy of quark/lepton masses 
and mixing angles appear, etc.
Non-abelian discrete symmetries are interesting 
ideas to address the above flavor issue.

It is plausible that such non-abelian discrete 
flavor symmetries are originated from extra dimensional 
theories, because non-abelian symmetries 
are symmetries of geometrical solids.
Indeed, in Ref.~\cite{Kobayashi:2004ya,Kobayashi:2006wq,Ko:2007dz} 
it has been shown that certain types 
of non-abelian discrete flavor symmetries 
such as $D_4$ and $\Delta(54)$ can appear 
in four-dimensional effective field theories 
derived from heterotic string theory with 
orbifold background.
(See also \cite{Altarelli:2006kg}.)
In those analyses, 
the important ingredients to derive 
the non-abelian discrete flavor symmetry 
are geometrical symmetries of the compact space 
and stringy coupling selection rules.
Thus, stringy non-abelian discrete flavor symmetries 
are, in general, larger than geometrical symmetries 
of the compact space.

It is important to extend such an analysis on 
heterotic orbifold models to 
other types of string models.
In this paper, we study which types of 
non-abelian flavor symmetries can appear from 
magnetized/intersecting brane models.
Magnetized D-brane models and intersecting 
D-brane models are T-duals of each other \cite{Bachas:1995ik}.\footnote{
See for a review \cite{Blumenhagen:2005mu} and references therein.}
Selection rules for allowed couplings 
in these models have been studied  \cite{Cremades:2003qj,Higaki:2005ie,
Cremades:2004wa,Abe:2009dr}.
Furthermore, three-point and higher order couplings 
have been computed explicitly~\cite{Cvetic:2003ch,Cremades:2004wa,DiVecchia:2008tm,Abe:2009dr,Russo:2007tc}.\footnote{See for three-point and 
higher order couplings and their selection rules in heterotic 
orbifold models~\cite{Hamidi:1986vh,Kobayashi:1991rp,Kobayashi:2004ya,
Buchmuller:2006ik}.}
Using these results, we study the flavor structures, 
which can appear in four-dimensional effective 
field theory derived from magnetized/intersecting 
brane models.
For concreteness, we study the flavor structure 
in magnetized brane models as well as magnetized 
orbifold models.
Then, we show several non-abelian discrete flavor symmetries 
can appear in magnetized brane models 
and they include $D_4$, $\Delta(27)$ and $\Delta(54)$, 
although $\Delta(27)$ is not realized in heterotic orbifold
models.\footnote{Indeed, these flavor symmetries are 
interesting for phenomenological model building.
See e.g. \cite{Grimus,Branco:1983tn,Ishimori:2008uc}.
}
Furthermore, most of their representations can appear 
in magnetized brane models, 
while certain representations appear in heterotic orbifold models.
We would obtain the same results in intersecting D-brane models, 
because of the T-duality between magnetized and intersecting 
D-brane models.

This paper is organized as follows.
In section 2, we review on magnetized brane models, 
in particular their zero-modes.
In section 3, we study three-point and higher 
order couplings and their selection rules.
In section 4, we study non-abelian discrete flavor 
symmetries, which can appear in magnetized brane models 
with non-vanishing Wilson lines.
Such analysis is extended to the models with vanishing 
Wilson lines in section 5 and enhancement of symmetries are shown.
In section 6, we discuss the flavor symmetries 
on the orbifold background.
Section 7 is devoted to conclusion and discussion.

\section{Magnetized brane models}

We start with ${\cal N}=1$ ten-dimensional $U(N)$ super Yang-Mills theory.
We consider the background $R^{3,1}\times (T^2)^3$,  
whose coordinates are denoted by
$x_\mu$ $(\mu=0,\cdots, 3)$ for the uncompact space $R^{3,1}$
and $y_m$ $(m=4, \cdots, 9)$ for the compact space $(T^2)^3$.
The Lagrangian is given by 
\begin{eqnarray}
{\cal L} &=& 
-\frac{1}{4g^2}{\rm Tr}\left( F^{MN}F_{MN}  \right) 
+\frac{i}{2g^2}{\rm Tr}\left(  \bar \lambda \Gamma^M D_M \lambda
\right),
\nonumber
\end{eqnarray}
where $M,N=0,\cdots, 9$.
Here, $\lambda$ denotes gaugino fields, $\Gamma^M$ is the 
gamma matrix for ten-dimensions and 
the covariant derivative $D_M$ is given as 
\begin{eqnarray}
D_M\lambda &=& \partial_M \lambda - i [A_M, \lambda],
\end{eqnarray}
where $A_M$ is the vector field.
Furthermore, the field strength $F_{MN}$ is given by 
\begin{eqnarray}
F_{MN} &=& \partial_M A_N - \partial_N A_M -i[A_M,A_N].
\end{eqnarray}
The gaugino fields $\lambda$ and the vector fields $A_m$ 
corresponding to the compact directions are decomposed as 
\begin{eqnarray}
\lambda(x,y) &=& \sum_n \chi_n(x) \otimes \psi_n(y), 
\nonumber \\
A_m(x,y) &=& \sum_n \varphi_{n,m}(x) \otimes \phi_{n,m}(y).
\nonumber
\end{eqnarray}

We factorize the six-torus into two-tori  $(T^2)^3$, each of which is
specified by the complex structure $\tau_d$ and the area 
$A_d = (2 \pi R_d)^2~\Im \tau_d$ where $d=1,2,3$.
We introduce the following form of the magnetic flux,
\begin{equation} \label{toronbg} \begin{split}
 F_{z^d \bar z^d} = {2 \pi \over \Im \tau_d}  \begin{pmatrix}
   m_1^{(d)} \1_{N_1} & & \\
  & \ddots & \\ & & m_n^{(d)} \1_{N_n} \end{pmatrix}, \quad d=1,2,3,
\end{split} \end{equation}
where $\1_{N_a}$ are the unit matrices of rank $N_a$, $m_i^{(d)}$ are
integers and we use the complex coordinates $z^d$. 
This background  breaks the gauge symmetry 
$U(N) \to \prod_{a=1}^n U(N_a)$ where
$N=\sum_{a=1}^n N_a.$

By introducing magnetic fluxes, we can realize four-dimensional 
chiral theory.
Let us focus on a submatrix consisting of two blocks,
\begin{equation}
 F_{z^d \bar z^d,ab} = {2\pi \over \Im \tau_d}
  \begin{pmatrix} m_a^{(d)} \1_{N_a} & 0 \\ 0 & m_b^{(d)} \1_{N_b} 
\end{pmatrix} .
\end{equation}
Then, 
the corresponding internal components $\psi_n(z)$ of gaugino fields 
$\lambda(x,z)$ also have the following form
\begin{equation} \label{KKdecomp}
 \psi_n(z) = \begin{pmatrix} \psi_n^{aa}(z) & \psi_n^{ab}(z) \\ \psi_n^{ba}(z) &
   \psi_n^{bb}(z) \end{pmatrix}.
\end{equation}
The off-diagonal components of zero-modes transform
as bifundamental representations
$\psi^{ab} \sim  \bf (N_a,\overline N_b)$, $\psi^{ba} \sim \bf
(\overline N_a, N_b)$
under $SU(N_a) \times SU(N_b)$, where we omit the subscript 0  
corresponding to the zero-modes, $n=0$.
For a fixed four-dimensional chirality, either $\psi^{ab}$ or 
$\psi^{ba}$ appears as zero-modes with normalizable wavefunctions, 
since the ten-dimensional chirality of $\lambda$ is fixed.
Which zero-modes appear,  $\psi^{ab}$ or 
$\psi^{ba}$, depends on the sign of the relative magnetic flux $M^{(d)}
\equiv m_a^{(d)}-m_b^{(d)}$.
Furthermore, the internal part $\psi(z)$ is decomposed as 
a product of the $d$-th $T^2$ part, i.e. $\psi_{(d)}(z^d)$, 
and each of them is two-component spinor.

With an appropriate gauge fixing, the zero-modes on
each $d$-th $T^2$ are written as \cite{Cremades:2004wa}
\begin{equation} \label{wavefn}
 \psi_d^{j,M^{(d)}}(z^{d}) = N_{M^{(d)}} 
~e^{i \pi M^{(d)} z^{d}{\Im~z^{d}/
   (\Im~\tau_d})} 
~\jtheta{j/M^{(d)} \\ 0}(M^{(d)}z^{d},\tau_dM^{(d)}),
\end{equation}
for $j=1,\dots, |M^{(d)}|$, where the normalization factor $N_M$ is obtained 
as
\begin{equation} \label{normalization}
N_{M} = \left( {2 \Im \tau_{d} |M| \over
   A_{d}^2 } \right)^{1/4} ,
\end{equation}
and $\jtheta{j/M^{(d)} \\ 0}(M^{(d)}z^{d},\tau_dM^{(d)})$ 
denotes the Jacobi theta function
\begin{equation} \label{jacobitheta}
 \jtheta{a \\ b}(\nu,\tau) = \sum_{n=-\infty}^\infty \exp\left[ \pi i
   (n+a)^2 \tau + 2 \pi i (n+a)(\nu +b)\right].
\end{equation}
We have the $|M^{(d)}|$ zero-modes labelled by the index $j$.
Note that the wavefunction for $j=k+M^{(d)}$ is identical to one 
for $j=k$.
The total number of zero-modes is the product, 
$\prod_d |M^{(d)}|$ and their wavefunctions are also 
given as the product, $\prod_d \psi_{(d)}^{j_d,M^{(d)}}$.
Furthermore, their flavor structure is also understood as 
a direct product of the $d$-th $T^2$ sector.
Thus, we concentrate on the $d$-th $T^2$ part 
and hereafter we omit the subscript $d$.
In addition, the relative magnetic flux $M$ is 
more important than the magnetic fluxes themselves, $m_a$ 
and $m_b$, from the viewpoint of the flavor structure.
Hence, we examine relative magnetic fluxes 
without mentioning the magnetic fluxes themselves, $m_a$ 
and $m_b$.

We can have Wilson lines, $\zeta \equiv \zeta_r + \tau \zeta_i$,
whose effect is just a translation of each wavefunction~\cite{Cremades:2004wa}
\begin{equation}
 \psi^{j,M}(z) \to \psi^{j,M}(z+\zeta), 
\end{equation}
for all of $j$.

\section{Coupling selection rule}

We study order $L$ couplings including the three point 
couplings $L=3$ in four-dimensional effective theory, i.e., 
\begin{equation}
 Y_{i_1 \dots i_{L_\chi} i_{L_\chi+1}\cdots i_L} \chi^{i_1}(x) \cdots 
\chi^{i_{L_\chi}}(x) \phi^{i_{L_{\chi}+1}}(x) \dots \phi^{i_L}(x),
\end{equation}
with $L=L_\chi + L_\phi$, where $\chi$ and $\phi$ collectively 
represent four-dimensional components of fermions and bosons,
respectively.
In particular, the selection rule for allowed couplings 
is important.
The three-point couplings can appear from the dimensional 
reduction of ten-dimensional super-Yang--Mills theory 
and higher order coupling terms can be 
read off from the effective Lagrangian of 
the Dirac--Born--Infeld action with
supersymmetrization. 
The internal component of bosonic and 
fermionic wavefunctions is the same \cite{Cremades:2004wa}.
Thus, the couplings are determined by the wavefunction overlap in the extra
dimensions,
\begin{equation}
 Y_{i_1 i_2 \dots i_L} = g_L^{10}  \int_{T^6} d^6z \ 
\prod_{d=1}^3 \psi^{i_1,M_1}_d(z) 
\psi^{i_2,M_2}_d(z) \dots \psi^{i_L,M_L}_d(z),
\end{equation}
where $g_L^{10}$ denotes the coupling in ten dimensions.
Here, as mentioned in the previous section, 
we concentrate on the two-dimensional $T^2$ part 
of the overlap integral of wavefunctions,
\begin{equation}
 y_{i_1 i_2 \dots i_L} = \int_{T^2} d^2z \ 
\psi^{i_1,M_1} (z) 
\psi^{i_2,M_2} (z) \dots \psi^{i_L,M_L}(z),
\end{equation}
where we have omitted the subscript $d$, again.

For example, we calculate the three-point couplings,
\begin{equation}
  y_{i_1i_2\bar i_3} = \int d^2 z \ 
\psi^{i_1,M_1}(z) \psi^{i_2,M_2}(z) \left( \psi^{i_3,M_3}(z) \right)^* .
\end{equation}
For the moment, we consider the case with vanishing Wilson lines.
The gauge invariance requires that $M_1+M_2 =M_3$ and 
that the wave function $\left( \psi^{i_3,M_3}(z) \right)^*$ but not 
$\psi^{i_3,M_3}(z)$ appears in the allowed three-point couplings.
If these are not satisfied, 
there is not corresponding operators in the ten dimensions, 
i.e. $g^{10}_3=0$.
The results are obtained as~\cite{Cremades:2004wa}
\begin{equation}\label{yijk-1}
 y_{i_1 i_2\bar i_3} = \sum_{m \in Z_{M_3}}  \delta_{i_1+i_2+M_1m,i_3}
 ~\jtheta{{M_2 i_1 - M_1 i_2 + M_1 M_2 m \over M_1 M_2 M_3}
 \\ 0}(0,\tau M_1 M_2 M_3),
\end{equation}
where the numbers in the Kronecker delta is defined modulo $M_3$. 
Indeed, the Kronecker delta part leads to the selection rule 
for allowed couplings as 
\begin{equation} \label{deltaconstraint}
 i_1+i_2-i_3 = M_3 l - M_1 m, \quad m \in { Z}_{M_3}, \ l \in {
   Z}_{M_1}.
\end{equation}
When $\gcd(M_1,M_2,M_3)=1$, every combination $(i_1,i_2,i_3)$ 
satisfies this constraint (\ref{deltaconstraint}) 
because of Euclidean algorithm.
On the other hand, when $\gcd(M_1,M_2,M_3)=g$, 
the above constraint becomes
\begin{equation} \label{deltaconstraint-g}
 i_1+i_2-i_3 = 0  \qquad (~{\rm mod} \ g~).
\end{equation}
This implies that we can define $Z_g$ charges from 
$i_k$ 
for zero-modes and the allowed couplings are controlled by 
such $Z_g$ symmetry.
Indeed, each quantum number $i_k$ corresponds to 
quantized momentum defined with the $M_i$ modulo structure.
When $\gcd(M_1,M_2,M_3)=g$, the modulo structure 
becomes $Z_g$ and the conservation law of 
these discrete momenta corresponds to a
requirement due to the $Z_g$ invariance.

Let us consider higher order couplings.
In~\cite{Abe:2009dr}, it has been shown that 
higher order couplings can 
be decomposed as productions of three-point couplings.
For example, we consider the four-point coupling,
\begin{equation}
  y_{i_1i_2i_3\bar i_4} = \int d^2 z \ 
\psi^{i_1,M_1}(z) \psi^{i_2,M_2}(z) \psi^{i_3,M_3}(z) 
\left( \psi^{i_4,M_4}(z) \right)^* .
\end{equation}
This four-point coupling can be decomposed as 
\begin{equation}
y_{i_1i_2i_3\bar i_4} = \sum_{s \in Z_M} y_{i_1 i_2 \bar s} 
\ y_{s i_3 \bar i_4},
\end{equation}
where 
\begin{eqnarray}
  y_{i_1i_2\bar s} &=& \int d^2 z \ 
\psi^{i_1,M_1}(z) \psi^{i_2,M_2}(z) \left( \psi^{s,M}(z)
\right)^*, \nonumber \\
 y_{s i_3\bar i_4} &=& \int d^2 z \ 
\psi^{s,M}(z) \psi^{i_3,M_3}(z) \left( \psi^{i_4,M_4}(z)
\right)^*,
\end{eqnarray}
with $M=M_1+M_2=M_4-M_3$.
Here, $\psi^{s,M}(z)$ denotes the $s$-th zero-mode of 
Dirac equation with the relative magnetic flux $M$, 
and these modes correspond to intermediate states 
in the above decomposition.
Each of $y_{i_1 i_2 \bar s} $ and $y_{s i_3 \bar i_4}$ 
is obtained as eq.~(\ref{yijk-1}).
That is, the coupling selection rule is controlled by 
the $Z_g$ invariance (\ref{deltaconstraint}), 
i.e. the conservation law of discrete momenta, 
and its modulo structure is determined by 
$\gcd(M_1,M_2,M_3,M_4)=g$.

Similarly, higher order couplings are decomposed 
as products of three-point couplings~\cite{Abe:2009dr}.
Therefore, the above analysis is generalized to 
generic order $L$ couplings.
That is, the coupling selection rule is 
given as the $Z_g$ invariance and its 
modulo structure is determined by 
$\gcd(M_1,\cdots ,M_L)=g$.

So far, we have considered the model with 
vanishing Wilson lines.
Non-vanishing Wilson lines do not affect 
the coupling selection rule due to the $Z_g$ 
invariance, but change 
values of couplings $y_{i_1 i_2 \bar i_3}$.
For example, when we introduce Wilson lines 
$\zeta_k$ for $\psi^{i_k,M_k}(z)$, 
the three-point coupling (\ref{yijk-1}) becomes  
\begin{eqnarray}\label{yijk-wl}
 y_{i_1 i_2\bar i_3} &=& \sum_{m \in \Z_{M_3}}
 \delta_{i_1+i_2+M_1m,i_3} e^{i\pi 
(\sum_{k=1}^3 M_k \zeta_k \Im \zeta_k)/\Im  \tau} \nonumber \\
& & \times 
 ~\jtheta{{M_2 i_1 - M_1 i_2 + M_1 M_2 m \over M_1 M_2 M_3}
 \\ 0}(M_2 M_3 (\zeta_2 - \zeta_3),\tau M_1 M_2 M_3),
\end{eqnarray}
where Wilson lines must satisfy $\zeta_3 M_3 = \zeta_1 M_1 + \zeta_2
M_2$.
Similarly, higher order couplings with non-vanishing 
Wilson lines can be obtained.

\section{Non-abelian flavor symmetries}

Here we study non-abelian flavor symmetries,
by using the analysis on the coupling selection rule 
in the previous section.

\subsection{Generic case}

First we study generic case with non-vanishing Wilson lines.
We consider the model with zero-modes $\psi^{i_k,M_k}$ 
for $k=1,\cdots, L$.
We denote $\gcd (M_1,\cdots,M_L) =g$.
As studied in the previous section, 
these modes have $Z_g$ charges and their 
couplings are controlled by the $Z_g$ invariance.
For simplicity, suppose that $M_1=g$.
Then, there are $g$ zero-modes of  $\psi^{i_1,M_1}$.
The above $Z_g$ transformation acts on $\psi^{i_1,g}$ as 
$Z \psi^{i_1,g}$, where
\begin{eqnarray}
Z = \left(
\begin{array}{ccccc}
1 & & & & \\
  & \rho & & & \\
  & & \rho^2 & & \\
  & &   & \ddots & \\
  &  &  &    & \rho^{g-1} 
\end{array}
\right),
\end{eqnarray}
and $\rho = e^{2\pi i /g}$.

In addition to this $Z_g$ symmetry, 
the effective theory has another symmetry.
That is, the effective theory must be 
invariant under cyclic permutations 
\begin{equation}\label{eq:permutation}
\psi^{i_1,g} \rightarrow \psi^{i_1+n,g} ,
\end{equation}
with a universal integer $n$ for $i_1$.
That is nothing but a change of ordering and also 
has a geometrical meaning as a discrete shift 
of the origin, $z=0 \rightarrow z=-\frac{n}{g}$.
This symmetry also generates another $Z_g$ symmetry, 
which we denote by $Z_g^{(C)}$ and its generator is 
represented as 
\begin{eqnarray}\label{eq:C}
C = \left(
\begin{array}{cccccc}
0 & 1& 0 & 0 & \cdots & 0 \\
0  & 0 &1 & 0 & \cdots & 0\\
  &    &  & &\ddots & \\
1  &  0  & 0 &  & \cdots   & 0 
\end{array}
\right),
\end{eqnarray}
on $\psi^{i_1,g}$.
That is, the above permutation (\ref{eq:permutation}) 
is represented as $C^n \psi^{i_1,g}$.
These generators, $Z$ and $C$, do not commute each other,
i.e., 
\begin{equation}
CZ = \rho ZC .
\end{equation}
Then, the flavor symmetry corresponds to the closed algebra 
including $Z$ and $C$.
Diagonal matrices in this closed algebra are written as 
$Z^n(Z')^m$, 
where $Z'$ is the generator of another $Z'_g$ and 
written as 
\begin{eqnarray}
Z' = \left(
\begin{array}{ccc}
\rho & &  \\
    & \ddots & \\
    &    & \rho 
\end{array}
\right),
\end{eqnarray}
on $\psi^{i_1,g}$.
Hence, these would generate the non-abelian flavor symmetry 
$(Z_g \times Z'_g)\rtimes Z_g^{(C)}$, since 
$Z_g \times Z'_g$ is a normal subgroup.
These discrte flavor groups would include $g^3$ elements totally.

Let us study actions of $Z$ and $C$ on 
other zero-modes, $\psi^{i_k,M_k}$, with $M_k=gn_k$, 
where $n_k$ is an integer.
First, the generator $C$ acts as 
\begin{equation}
\psi^{i,gn_k} \rightarrow \psi^{i+n_k,gn_k},
\end{equation}
because the above discrete shift of the origin 
$z=0 \rightarrow z=-\frac{n}{g}$ can be written as 
$z=0 \rightarrow z=-\frac{nn_k}{gn_k}$
for these zero-modes.
Thus, the generator $C$ is represented as the same as 
(\ref{eq:C}) on the basis
\begin{eqnarray}\label{eq:g-plet}
\left(
\begin{array}{c}
\psi^{p,gn_k}  \\ 
\psi^{p+n_k,gn_k}  \\ 
\vdots  \\
\psi^{p+(g-1)n_k,gn_k} 
\end{array}
\right),
\end{eqnarray}
where $p$ is an integer.
Note that $\psi^{p+gn_k,gn_k}$ is identical to $\psi^{p,gn_k}$.
Furthermore, the generator $Z$ is represented on this basis 
(\ref{eq:g-plet}) as 
\begin{eqnarray}
Z = \rho^p \left(
\begin{array}{ccccc}
1 & & & & \\
  & \rho^{n_k} & & & \\
   & & \rho^{2n_k} & & \\
  &    & & \ddots & \\
  &    & &   & \rho^{(g-1)n_k} 
\end{array}
\right).
\end{eqnarray}
Thus, the zero-modes $\psi^{i_k,gn_k}$ include $n_k$ $g$-plet 
representations of the symmetry $(Z_g \times Z'_g) \rtimes Z_g^{(C)}$ 
and some of them may be reducible $g$-plet 
representations.
For example, when we consider 
the zero-modes corresponding to $n_k=g$, i.e. $M_k=g^2$, 
the generator $Z$ is represented as 
$\rho^p {\1}_g$ on the above $g$-plet (\ref{eq:g-plet}), 
where ${\1}_g$ is the $(g \times g)$ unit matrix.
In such a case, the generator $C$ can also be diagonalized.
Then, these zero-modes correspond to $g$ singlets of 
$(Z_g \times Z'_g) \rtimes Z_g^{(C)}$ 
including trivial and non-trivial singlets.

As illustrating examples, we consider the models with 
$g=2,3$ in the next subsections and study 
more concretely about non-abelian discrete flavor symmetries.

\subsection{$g=2$ model}

Here we consider the model with $g=2$, that is, 
all of relative magnetic fluxes $M_k$ are even.
Its flavor symmetry is given as the closed algebra of 
$Z_2$, $Z'_2$ and $Z_2^{(C)}$, and all of these elements are 
written as 
\begin{eqnarray}
\pm \left(
\begin{array}{cc}
1 & 0 \\
0 & 1 \\
\end{array}
\right), \quad 
\pm \left(
\begin{array}{cc}
0 & 1 \\
1 & 0 \\
\end{array}
\right), \quad 
\pm \left(
\begin{array}{cc}
0 & 1 \\  -1 & 0 \\
\end{array}
\right), \quad 
\pm \left(
\begin{array}{cc}
1 & 0 \\
0 & -1 \\
\end{array}
\right).
\end{eqnarray}
That is,  the flavor symmetry is $D_4$.
The zero-modes with the relative magnetic flux 
$M=2$,
\begin{eqnarray}
\left(
\begin{array}{c}
\psi^{0,2} \\
\psi^{1,2}
\end{array}
\right),
\end{eqnarray}
correspond to the doublet representation ${\bf 2}$ of $D_4$.
This result is the same as the non-abelian flavor symmetry 
appearing from heterotic orbifold models with $S^1/Z_2$, 
where twisted modes on two fixed points of $S^1/Z_2$ 
correspond to the $D_4$ doublet~\cite{Kobayashi:2004ya,Kobayashi:2006wq}.

Next, we consider the zero-modes corresponding to 
the relative magnetic flux $M=4$, $\psi^{i,4}$ ($0=0,1,2,3$).
As discussed in the previous subsection, 
in order to represent $C$, 
it may be convenient to decompose them into the $g$-plets 
(\ref{eq:g-plet})
\begin{eqnarray}
\left(
\begin{array}{c}
\psi^{0,4} \\
\psi^{2,4}
\end{array}
\right), \qquad 
\left(
\begin{array}{c}
\psi^{1,4} \\
\psi^{3,4}
\end{array}
\right).
\end{eqnarray}
However, they are reducible representations as follows.
Note that both $\psi^{0,4}$ and $\psi^{2,4}$ have 
even $Z_2$ charges, and that both $\psi^{1,4}$ and $\psi^{3,4}$ have 
odd $Z_2$ charges.
That is, the generator $Z$ is represented in the form $\pm {\bf 1}_2$, where 
${\bf 1}_2$ is the $2 \times 2$ identity matrix.
Thus, the generator $C$ can be diagonalized and such a diagonalizing 
basis is obtained as
\begin{eqnarray}\label{eq:D4-singlets}
&&{\bf 1}_{++}:\ (\psi^{0,4} + \psi^{2,4}), \quad \
{\bf 1}_{+-}:\ (\psi^{0,4} - \psi^{2,4}), \quad \ \nonumber \\
&&{\bf 1}_{-+}:\ (\psi^{1,4} + \psi^{3,4}), \quad \
{\bf 1}_{--}:\ (\psi^{1,4} - \psi^{3,4}), \quad \
\end{eqnarray}
up to normalization factors.
Obviously, these correspond to four $D_4$ singlets, 
${\bf 1}_{++}$, ${\bf 1}_{+-}$, ${\bf 1}_{-+}$ 
and ${\bf 1}_{--}$.
The first subscript of two denotes $Z_2$ charges for $Z$ 
and the second one denotes $Z_2$ charges for $C$.
Hence, all of irreducible representations of $D_4$ appear from 
$\psi^{i,2}$ and $\psi^{i,4}$.
New representations can not appear in zero-modes 
$\psi^{i,M}$ with $M >4$.
For example,  we consider zero-modes corresponding to $M=6$, 
i.e. $\psi^{i,6}$.
They can be decomposed as 
\begin{equation}\label{eq:g2-6}
|\psi^{6}\rangle_1=
\left(
\begin{array}{c}
\psi^{0,6} \\ \psi^{3,6} 
\end{array} \right), \quad
|\psi^{6}\rangle_2=
\left(
\begin{array}{c}
\psi^{2,6} \\ \psi^{5,6} 
\end{array} \right), \quad
|\psi^{6}\rangle_3=
\left(
\begin{array}{c}
\psi^{4,6} \\ \psi^{1,6} 
\end{array} \right). \
\end{equation}
Each of $|\psi^{6}\rangle_i$ with $i=1,2,3$ is nothing but 
the $D_4$ doublet.
That is, we have three $D_4$ doublets in $\psi^{i,6}$.
The above representations appear repeatedly 
in $\psi^{i,M}$ with larger $M$.
These results are shown in Table~\ref{tab:g2}.

\begin{table}[t]
\begin{center}
\begin{tabular}{|c|c|} \hline
$M$ & Representation of $D_4$ \\ \hline \hline
2 & ${\bf 2}$ \\ 
4 & ${\bf 1}_{++}, \ {\bf 1}_{+-}, \ {\bf 1}_{-+}, \ {\bf 1}_{--}$ \\ 
6 & $3 \times {\bf 2}$\\  \hline
\end{tabular}
\end{center}
\caption{$D_4$ representations of zero-modes in the model with $g=2$.}
\label{tab:g2}
\end{table}

\subsection{$g=3$ model}

Here we consider the model with $g=3$, where 
all of relative magnetic fluxes are equal to $M_k=3n_k$.
Its flavor symmetry is given as $(Z_3 \times Z_3) \rtimes Z_3$, that is, 
$\Delta(27)$~\cite{Branco:1983tn}.
This flavor symmetry is different from the flavor symmetry 
appearing from heterotic orbifold models with $T^2/Z_3$.
Later, we will explain what makes this difference.

The zero-modes corresponding to the relative magnetic flux $M=3$,
\begin{eqnarray}\label{eq:g3-3}
|\psi^{3}\rangle_1=\left(
\begin{array}{c}
\psi^{0,3} \\
\psi^{1,3}  \\
\psi^{2,3} 
\end{array}
\right),
\end{eqnarray}
correspond to the triplet representation ${\bf 3}$ of $\Delta(27)$.
Next, we consider the zero-modes corresponding to 
the relative magnetic flux $M=6$, i.e. $\psi^{i,6}$.
Again, it may be convenient to decompose 
them into the $g$-plets (\ref{eq:g-plet})
\begin{equation}\label{eq:g3-6}
|\psi^{6}\rangle_1=
\left(
\begin{array}{c}
\psi^{0,6} \\ \psi^{2,6} \\ \psi^{4,6} 
\end{array} \right), \quad
|\psi^{6}\rangle_2=
\left(
\begin{array}{c}
\psi^{3,6} \\ \psi^{5,6} \\ \psi^{1,6} 
\end{array} \right).
\end{equation} 
The generator $C$ is represented in the same way for 
$|\psi^{3}\rangle_1$ and $|\psi^{6}\rangle_i$ ($i=1,2$).
On the other hand, the representation of the generator $Z$ 
for $|\psi^{6}\rangle_i$ ($i=1,2$) is the complex conjugate 
to one for $|\psi^{3}\rangle_1$.
Thus, both $|\psi^{6}\rangle_i$  ($i=1,2$) correspond to 
$\bar {\bf 3}$ representations of $\Delta(27)$.

Moreover, let us consider the zero-modes with the relative 
magnetic flux $M=9$, i.e. $\psi^{i,9}$.
Then, we decompose 
them into the $g$-plets (\ref{eq:g-plet})
\begin{equation}
|\psi^{9}\rangle_1=
\left(
\begin{array}{c}
\psi^{0,9} \\ \psi^{3,9} \\ \psi^{6,9} 
\end{array} \right), \quad
|\psi^{9}\rangle_\omega=
\left(
\begin{array}{c}
\psi^{1,9} \\ \psi^{4,9} \\ \psi^{7,9} 
\end{array} \right), \quad
|\psi^{9}\rangle_{\omega^2}=
\left(
\begin{array}{c}
\psi^{2,9} \\ \psi^{5,9} \\ \psi^{8,9} 
\end{array} \right),
\end{equation} 
where $\omega = e^{2\pi i/3}$.
These (reducible) triplets $ |\psi^{9}\rangle_{\omega^n}$ have 
$Z_3$ charges, $n$ and are decomposed into 
nine singlets,
\begin{equation}\label{eq:g3-9singlets}
{\bf 1}_{\omega^n,\omega^m}:\psi^{n,9}+\omega^m\psi^{n+3m,9}
+\omega^{2m} \psi^{n+6m,9},
\end{equation}
up to normalization factors, where 
$n$ and $m$ are $Z_3$ charges for $Z$ and $C$, respectively.
In zero-modes with $M >9$, new representations do not appear, 
but the above representations appear repeatedly.
These results as well as zero-modes with $M >9$ 
are shown in Table~\ref{tab:g3}.
Similar analysis can be carried out in other models 
with $g > 3$.

\begin{table}[t]
\begin{center}
\begin{tabular}{|c|c|} \hline
$M$ & Representation of $\Delta(27)$ \\ \hline \hline
3 & ${\bf 3}$ \\ 
6 & $2 \times {\bar {\bf 3}}$  \\ 
9 & ${\bf 1}_{1}, \ {\bf 1}_{2}, \ {\bf 1}_{3}, \ {\bf 1}_{4}, 
\ {\bf 1}_{5}, \ {\bf 1}_{6}, \ {\bf 1}_{7}, \ {\bf 1}_{8}, 
\ {\bf 1}_{9}$ \\ 
12 & $4 \times {\bf 3}$\\  
15 & $5 \times {\bar {\bf 3}}$  \\ 
18 & $2 \times \{ {\bf 1}_{1}, \ {\bf 1}_{2}, \ {\bf 1}_{3}, \ {\bf 1}_{4}, 
\ {\bf 1}_{5}, \ {\bf 1}_{6}, \ {\bf 1}_{7}, \ {\bf 1}_{8}, 
\ {\bf 1}_{9} \}$ \\ \hline
\end{tabular}
\end{center}
\caption{$\Delta(27)$ representations of zero-modes in the model with $g=3$.}
\label{tab:g3}
\end{table}


We comment on symmetries in subsectors.
Suppose that our model has zero-modes $\psi^{i_k,M_k}$ for 
$k = 1,\cdots, L$ with $\gcd(M_1,\cdots,M_L)=g$ and 
they are separated into two classes, 
$\psi^{i_l,M_l}$ $(l=1,\cdots,L_1)$ and 
$\psi^{i_m,M_m}$ $(m=L_1,\cdots,L)$, 
where $\gcd(M_1,\cdots,M_{L_1})=g_1$,  
$\gcd(M_{L_1},\cdots,M_{L})=g_2$ and 
$\gcd (g_1,g_2)=g$.
Coupling terms  including only the first class of 
fields $\psi^{i_l,M_l}$ $(l=1,\cdots,L_1)$  
in the four-dimensional effective theory have the 
symmetry $(Z_{g_1} \times Z_{g_1})\rtimes Z_{g_1}$, where 
$g_1$ would be larger than $g$.
However, such a symmetry is broken 
by terms including the second class of fields.
Thus, we would have a larger symmetry 
at least  at tree level for the subsectors.
Such larger symmetries in the subsectors would be interesting 
for model building.

\section{Models without Wilson lines}

In the previous section, we have considered the 
models with non-vanishing Wilson lines.
Here, we study the models without Wilson lines.
In this case, flavor symmetries are enhanced.

When Wilson lines are vanishing, all of zero-modes
$\psi^{0,M_k}$ have the peak at the same point 
in the extra dimensions.
In the intersecting D-brane picture, 
this corresponds to the D-brane configuration, that 
all of D-branes intersect (at least) at a single point on $T^2$.
This model has the $Z_2$ rotation symmetry around such a point.
Here, we denote its generator as $P$.
In general, this acts as 
\begin{equation}
P: \ \psi^{i,M} \rightarrow \psi^{M-i,M}.
\end{equation}

As in the previous section, we consider the models with 
$g=2,3$ as illustrating models.

\subsection{$g=2$ model}

First, we consider the zero-modes with $M=2$, 
$\psi^{i,2}$, which correspond to the $D_4$ doublet.
For them, the generator $P$ acts as the identity.
That implies that the flavor symmetry is enhanced as 
$D_4 \times Z_2$ and $\psi^{i,2}$ correspond to 
${\bf 2}_+$, where the subscript denotes 
the $Z_2$ charge for $P$.\footnote{
Although this is just an enhancement by the factor $Z_2$, 
such an enhanced flavor symmetry $D_4 \times Z_2$ 
would be important to phenomenological model building.
See e.g.~\cite{Grimus}.}

We consider the zero-modes with $M=4$, 
$\psi^{i,4}$, which are decomposed as 
the four $D_4$ singlets, 
${\bf 1}_{++}$, ${\bf 1}_{+-}$, ${\bf 1}_{-+}$ 
and ${\bf 1}_{--}$
as (\ref{eq:D4-singlets}).
They have definite $Z_2$ charges for $P$ and 
are represented as 
\begin{eqnarray}\label{eq:D4-singlets-2}
&&{\bf 1}_{+++}:\ (\psi^{0,4} + \psi^{2,4}), \quad \
{\bf 1}_{+-+}:\ (\psi^{0,4} - \psi^{2,4}), \quad \ \nonumber \\
&&{\bf 1}_{-++}:\ (\psi^{1,4} + \psi^{3,4}), \quad \
{\bf 1}_{---}:\ (\psi^{1,4} - \psi^{3,4}), \quad \
\end{eqnarray}
where the third sign in the subscripts 
denotes $Z_2$ charges for $P$.

Now, let us consider the zero-modes with $M=6$, 
$\psi^{i,6}$, which are decomposed as 
three $D_4$ doublets (\ref{eq:g2-6}).
The doublet $|\psi^{6}\rangle_1$ has the even 
$Z_2$ charges for $P$.
However, other doublets $|\psi^{6}\rangle_2$ and 
$|\psi^{6}\rangle_3$ transform each other under $P$.
Thus, we take linear combinations of these two doublets as
\begin{eqnarray}
|\psi^{6}\rangle_\pm \equiv 
|\psi^{6}\rangle_2 \pm
|\psi^{6}\rangle_3 
=
\left(
\begin{array}{c}
\psi^{2,6} \\ \psi^{5,6} 
\end{array} \right) \pm
\left(
\begin{array}{c}
\psi^{4,6} \\ \psi^{1,6} 
\end{array} \right) ,
\end{eqnarray}
where $\pm$ also means $Z_2$ charge of $P$.
As a result, these zero-modes $\psi^{i,6}$ are 
decomposed as two ${\bf 2}_+$ and one ${\bf 2}_-$.

We can repeat these analysis for larger $M$.
For example, zero-modes with $M=8$, 
$\psi^{i,8}$, are decomposed as 
\begin{equation}
\{{\bf 1}_{+++},\ {\bf 1}_{+-+},\ 
           {\bf 1}_{+++},\ {\bf 1}_{+--},\
           {\bf 1}_{-++},\ {\bf 1}_{-+-},\ 
           {\bf 1}_{---},\ {\bf 1}_{--+}\} , 
\end{equation}
and zero-modes with $M=10$, 
$\psi^{i,10}$, are decomposed as 
three ${\bf 2}_+$ and two ${\bf 2}_-$.
These results are shown in Table~\ref{tab:g2-2}.

\begin{table}[t]
\begin{center}
\begin{tabular}{|c|c|} \hline
$M$ & Representation of $D_4\times Z_2$ \\ \hline \hline
2 & ${\bf 2}_+$ \\ 
4 & ${\bf 1}_{+++}, \ {\bf 1}_{+-+}, \ {\bf 1}_{-++}, \ {\bf 1}_{---}$ \\ 
6 & $2 \times {\bf 2}_+, \ {\bf 2}_-$\\  
8 & ${\bf 1}_{+++},\ {\bf 1}_{+-+},\ 
           {\bf 1}_{+++},\ {\bf 1}_{+--},\
           {\bf 1}_{-++},\ {\bf 1}_{-+-},\ 
           {\bf 1}_{---},\ {\bf 1}_{--+}$ \\ 
10 & $3 \times {\bf 2}_+, \ 2 \times {\bf 2}_-$\\  \hline
\end{tabular}
\end{center}
\caption{$D_4 \times Z_2$ representations of zero-modes in the model with $g=2$.}
\label{tab:g2-2}
\end{table}

\subsection{$g=3$ model}

Here, we study the model with $g=3$.
First, we consider the zero-modes with $M=3$, $\psi^{i,3}$.
They correspond to a triplet of $\Delta(27)$ with non-vanishing 
Wilson lines.
At any rate, the generators, $Z$, $C$ and $P$, acts on $\psi^{i,3}$ as 
\begin{eqnarray}\label{eq:delta-54}
Z=\left(
\begin{array}{ccc}
1 & 0 & 0 \\
0 & \omega & 0 \\
0 & 0 & \omega^2 
\end{array}
\right), \quad
C = \left(
\begin{array}{ccc}
0 & 1 & 0 \\
0 & 0 & 1 \\
1 & 0 & 0
\end{array}
\right), \quad
P = \left(
\begin{array}{ccc}
1 &  0 & 0 \\
0 & 0 & 1 \\
0 & 1 & 0
\end{array}
\right).
\end{eqnarray}
Their closed algebra is $\Delta(54)$.
Thus, the zero-modes $\psi^{i,3}$ correspond to 
the triplet of $\Delta(54)$, ${\bf 3}_1$.
This is the same as the flavor symmetry, which appears in
heterotic orbifold models with $T^2/Z_3$~\cite{Kobayashi:2006wq}.
Three fixed points on the orbifold $T^2/Z_3$ have the geometrical 
permutation symmetry $S_3$.
Such symmetry is enhanced in magnetized brane models, 
only when Wilson lines are vanishing.
Indeed, the closed algebra of generators $C$ and $P$ 
is $S_3$.

Similarly, we can consider the zero-modes with $M=6$, $\psi^{i,6}$.
We decompose them as (\ref{eq:g3-6}).
The generators, $C$ and $P$, act on 
$|\psi^{6}\rangle_i$ ($i=1,2$) in the same way as 
$\psi^{i,3}$, but the representation of the generator $Z$ 
for $|\psi^{6}\rangle_i$ ($i=1,2$) is the complex conjugate 
to one for $|\psi^{3}\rangle_1$.
Thus, both $|\psi^{6}\rangle_i$ correspond to 
$\bar {\bf 3}_1$ representations of $\Delta(54)$.
Recall that $|\psi^{6}\rangle_i$ are 
$\bar {\bf 3}$ representations of $\Delta(27)$.

Next, let us consider the zero-modes with $M=9$, $\psi^{i,9}$.
Recall that they correspond to nine singlets of $\Delta(27)$ as 
(\ref{eq:g3-9singlets}).
The following linear combination,
\begin{equation}
\psi^{0,9}+\psi^{3,9}+\psi^{6,9},
\end{equation}
is still a singlet under $\Delta(54)$, which is 
a trivial singlet ${\bf 1}_1$.
However, the others in linear combinations (\ref{eq:g3-9singlets}) 
transform each other under $P$.
Then, they correspond to four doublets of $\Delta(54)$,
\begin{eqnarray}
\begin{array}{cccc}
{\bf 2}_1 :&
\left(
\begin{array}{c}
\psi^{0,9}+\omega \psi^{3,9}+\omega^2 \psi^{6,9} \\
\psi^{0,9}+\omega^2 \psi^{3,9}+\omega \psi^{6,9}  
\end{array} 
\right), \quad &
{\bf 2}_2 :&
\left(
\begin{array}{c}
\psi^{1,9}+\psi^{4,9}+ \psi^{7,9} \\
\psi^{2,9}+\psi^{5,9}+ \psi^{8,9}  
\end{array} 
\right), \\
 & & & \\
{\bf 2}_3 :&
\left(
\begin{array}{c}
\psi^{1,9}+\omega \psi^{4,9}+\omega^2 \psi^{7,9} \\
\psi^{8,9}+\omega^2 \psi^{5,9}+\omega^2 \psi^{2,9}  
\end{array} 
\right),  \quad &
{\bf 2}_4 :&
\left(
\begin{array}{c}
\psi^{1,9}+\omega^2 \psi^{4,9}+\omega \psi^{7,9} \\
\psi^{8,9}+\omega^2 \psi^{5,9}+\omega \psi^{2,9}  
\end{array} 
\right).
\end{array} 
\end{eqnarray}

Now, let us consider the zero-modes with $M=12$, $\psi^{i,12}$.
We decompose them into $g$-plets 
(\ref{eq:g-plet})
\begin{eqnarray}
 & & |\psi^{12} \rangle_1 =
\left(
\begin{array}{c}
\psi^{0,12} \\ \psi^{4,12} \\ \psi^{8,12} 
\end{array} \right), \qquad
|\psi^{12} \rangle_2 =
\left(
\begin{array}{c}
\psi^{6,12} \\ \psi^{10,12} \\ \psi^{2,12} 
\end{array} \right), \nonumber\\ 
 & & 
|\psi^{12} \rangle_3 =
\left(
\begin{array}{c}
\psi^{3,12} \\ \psi^{7,12} \\ 
\psi^{11,12}
\end{array} \right), \qquad
|\psi^{12} \rangle_4 =
\left(
\begin{array}{c}
\psi^{9,12} \\ \psi^{1,12} \\ 
\psi^{5,12} 
\end{array} \right). 
\end{eqnarray}
They correspond to four triplets of $\Delta(27)$.
Representations of the generators, $Z$, $C$ and $P$, 
on $|\psi^{12} \rangle_1$ and  $|\psi^{12} \rangle_2$ 
are the same as those on  $\psi^{i,3}$ like 
Eq.~(\ref{eq:delta-54}).
Thus, they correspond to ${\bf 3}_1$.
On the other hand, $|\psi^{12} \rangle_3$ and  $|\psi^{12} \rangle_4$ 
transform each other under $P$.
Hence, we take the following linear combinations,
\begin{eqnarray}
& &  |\psi^{12} \rangle_{\pm} =
\left(
\begin{array}{c}
\psi^{3,12}\pm \psi^{9,12} \\ \psi^{7,12}\pm \psi^{1,12} \\ 
\psi^{11,12}\pm \psi^{5,12} 
\end{array} \right).
\end{eqnarray}
Then, representations of $Z$, $C$ and $P$ on 
$|\psi^{12} \rangle_+$ are the same as (\ref{eq:delta-54}), 
and $|\psi^{12} \rangle_+$  corresponds to ${\bf 3}_1$.
On the other hand, representations of $Z$ and $C$ on 
$|\psi^{12} \rangle_-$ are the same as (\ref{eq:delta-54}), 
but the generator $P$ is represented on $|\psi^{12} \rangle_-$ as 
\begin{eqnarray}
P = \left(
\begin{array}{ccc}
 -1 &  0 & 0 \\
 0 & 0 & -1 \\
 0 & -1 & 0
\end{array}
\right).
\end{eqnarray}
That is, $|\psi^{12} \rangle_-$ corresponds to 
another triplet of $\Delta(54)$, i.e. ${\bf 3}_2$.
Furthermore, the zero-modes with $M=15$, $\psi^{i,15}$ 
correspond to 
\begin{eqnarray}
& & 3 \times \bar{{\bf 3}}_1,\quad 
           2 \times \bar{{\bf 3}}_2 ,
\end{eqnarray}
and the zero-modes with $M=18$, $\psi^{i,18}$ 
correspond to 
\begin{eqnarray}
& & 2 \times \{ {\bf 1}_1, \ {\bf 2}_1, \ {\bf 2}_2, \ {\bf 2}_3, \
{\bf 2}_4 \}  .
\end{eqnarray}
These results are shown in Table~\ref{tab:g3-2}.
Irreducible representations of $\Delta(54)$ are 
two triplets ${\bf 3}_1$, ${\bf 3}_2$, their conjugates 
${\bar {\bf 3}}_1$ ${\bar {\bf 3}}_2$, four doublets 
${\bf 2}_1$, ${\bf 2}_2$, ${\bf 2}_3$, ${\bf 2}_4$, 
trivial singlet ${\bf 1}$ and non-trivial singlet ${\bf 1}_2$.
All of them except the non-trivial singlet ${\bf 1}_2$ 
can appear in this model.

\begin{table}[t]
\begin{center}
\begin{tabular}{|c|c|} \hline
$M$ & Representation of $\Delta(54)$ \\ \hline \hline
3 & ${\bf 3}_1$ \\ 
6 & $2 \times {\bar {\bf 3}}_1$  \\ 
9 & ${\bf 1}_1, \ {\bf 2}_1, \ {\bf 2}_2, \ {\bf 2}_3, \ {\bf 2}_4$ \\ 
12 & $3 \times {\bf 3}_1, \ {\bf 3}_2$\\  
15 & $3 \times {\bar {\bf 3}}_1, \ 
2 \times {\bar {\bf 3}}_2$  \\ 
18 & $2 \times \{ {\bf 1}_1, \ {\bf 2}_1, \ {\bf 2}_2, \ {\bf 2}_3, \
{\bf 2}_4 \}$  \\
 \hline
\end{tabular}
\end{center}
\caption{$\Delta(54)$ representations of zero-modes in the model with $g=3$.}
\label{tab:g3-2}
\end{table}

Similar analysis can be carried out in other models with $g>3$.
In generic case, the $Z$ and $P$ satisfy 
\begin{equation}
PZ = Z^{-1}P,
\end{equation} 
and the closed algebra of $C$ and $P$ is 
$D_g$.
Thus, the flavor symmetry, which is generated by 
$Z$, $C$ and $P$, would be written as $D_g \ltimes (Z_g \times Z_g)$.
Note that $S_3 \sim D_3$ and 
$\Delta(54)$ is $D_3 \ltimes (Z_3 \times Z_3)$.

\section{Orbifold models}

We have found that several non-abelian discrete flavor symmetries 
like $D_4$, $\Delta(27)$ and $\Delta(54)$ can appear.
However, these exact symmetries may be rather large 
to explain realistic mass matrices of quarks and leptons.
Their breaking would be preferable.
Such symmetry breaking can happen within the framework 
of four-dimensional effective field theory, 
that is, 
scalar fields with non-trivial representations 
are assumed to develop their vacuum expectation values.
On the other hand, a certain type of symmetry breaking 
can happen on the orbifold background, which is 
called magnetized orbifold models~\cite{Abe:2008fi,Abe:2008sx}.
Here, we discuss the flavor structure in 
magnetized orbifold models.

The orbifold $T^2/Z_2$ is constructed by 
dividing $T^2$ by the $Z_2$ projection $z \rightarrow -z$.
Furthermore, on such an orbifold, we require periodic 
or anti-periodic boundary condition for matter fields 
as well as gauge fields,
\begin{eqnarray}\label{eq:orbifold-bc}
& & 
\psi (-z) = \pm  \ \psi (z).
\end{eqnarray}
Since such boundary conditions are consistent in models 
with vanishing Wilson lines, we consider 
the case without Wilson lines.
Indeed, zero-mode wavefunctions in models without 
Wilson lines satisfy the following relation,
\begin{eqnarray}
& & 
\psi^{j,M}(-z) = \psi^{M-j,M}(z).
\end{eqnarray}
Thus, even and odd zero-modes are obtained as 
their linear combinations, 
\begin{eqnarray}
& & 
\psi^j_{\pm}(z)  = \psi^{j,M}(z) \pm \psi^{M-j,M}(z),
\end{eqnarray}
up to a normalization factor.
Which modes among even and odd modes are selected 
depends on how to embed the $Z_2$ orbifold projection 
into the gauge space, 
that is, model dependent.
At any rate, either even or odd zero-modes are 
projected out for each kind of matter fields\footnote{Within 
the framework of intersecting D-brane
  models, analogous results have been obtained by considering 
D6-branes wrapping rigid 3-cycles~\cite{Blumenhagen:2005tn}.}.
Note that the $Z_2$ orbifold parity of $\psi^j_{\pm}(z) $ 
is the same as the $Z_2$ charge of $P$.
Thus, through the orbifold projection 
zero-modes with either even or odd $Z_2$ charge of $P$ 
survive for each kind of matter fields.

Let us consider examples.
First we study the model with $g=2$.
This model has the non-abelian flavor symmetry 
$D_4 \times Z_2$.
The zero-modes with $M=2$, $\psi^{i,2}$, correspond 
to ${\bf 2}_+$ of $D_4 \times Z_2$.
When we require the periodic boundary condition, 
they survive.
On the other hand, they are projected out 
for the anti-periodic boundary condition.
Similarly, the zero-modes with 
$M=4$, $\psi^{i,4}$, correspond 
to ${\bf 1}_{+++}$, ${\bf 1}_{+-+}$, ${\bf 1}_{-++}$ and 
${\bf 1}_{---}$, where the third subscript denotes 
the  $Z_2$ charge of $P$.
Thus, the zero-modes corresponding to 
 ${\bf 1}_{+++}$, ${\bf 1}_{+-+}$ and ${\bf 1}_{-++}$
survive for the periodic boundary condition, 
while only ${\bf 1}_{---}$ survives 
for the anti-periodic boundary condition.
Similarly, we can identify which modes can survive 
through the $Z_2$ orbifold projection.
The number of matter fields are reduced through 
the $Z_2$ orbifold projection.
However, four-dimensional effective field theory 
after orbifolding has the flavor symmetry $D_4 \times Z_2$.
The reason why the flavor symmetry $D_4 \times Z_2$ 
remains unbroken is that the flavor symmetry 
is the direct product between $D_4$ and $Z_2$.

Next, let us consider the model with $g=3$.
This model has the flavor symmetry $\Delta(54)$.
The zero-modes with $M=3$, $\psi^{i,3}$, correspond 
to ${\bf 3}_1$ of $\Delta(54)$.
However, the eigenstates of $Z_2$ are 
$\psi^{0,3}$ and $\psi^{1,3} \pm \psi^{2,3}$.
Hence, when we project out $Z_2$ even or odd modes, 
the triplet structure is broken, that is, 
the flavor symmetry $\Delta(54)$ is completely 
broken.
However, such symmetry breaking is non-trivial, 
because the original theory has the $\Delta(54)$ symmetry 
and we project out certain modes from such a theory.\footnote{
This type of flavor symmetry breaking has been proposed in
not magnetized brane models, but orbifold 
models~\cite{Haba:2006dz,Kobayashi:2008ih,Seidl:2008yf}.}

Orbifold models with larger $g$, $g > 3$ have a similar 
structure on flavor symmetries.
The original theory before orbifolding has 
a large non-abelian flavor symmetry.
By orbifolding, certain matter fields are projected out 
and the flavor symmetry is broken although some symmetries 
like abelian discrete symmetries remain unbroken.
However, there remains a footprint of the larger flavor symmetry
in four-dimensional effective theory, 
that is, coupling terms are constrained.

As an illustrating example, let us consider explicitly 
the model with three zero-modes, which have 
relative magnetic fluxes, $(M_1,M_2,M_3)=(4,4,8)$, 
that is, $g=4$.
The generators, $Z$, $C$ and $P$, are represented 
on the zero-modes with $M_1=4$ as 
\begin{equation}
Z=
\left(
\begin{array}{cccc}
1 & & & \\  & i & & \\ & & -1 & \\ & & & -i 
\end{array}
\right), \ \ 
C=
\left(
\begin{array}{cccc}
& 1 & & \\  & & 1 & \\ & & & 1  \\ 1 & & &  
\end{array}
\right), \ \ 
P=
\left(
\begin{array}{cccc}
1 & & & \\  &  & & 1 \\ & & 1 & \\ & 1 & &  
\end{array}
\right).
\end{equation}
Obviously, we find $[P,Z] \ne 0$ and $[C,P] \ne 0$.
Thus, eigenstates of $P$ are not eigenstates for 
$Z$ or $C$.
Since eigenstates with $P=1$ or $P=-1$ are 
projected out by orbifolding, the flavor symmetry 
is broken.
However, one can find that $[P,Z^2]=[P,C^2]=0$.
The symmetry generated by $Z^2$, $C^2$ and $P$ 
remains unbroken after orbifolding.
Thus, the flavor symmetry is reduced to 
$Z_2 \times Z_2 \times Z_2$.
The first two $Z_2$ factors are originally subgroups of 
$Z_4 \ltimes (Z_4 \times Z_4)$ generated by $Z$ and $C$ algebra and 
they are abelian groups.

For concreteness, let us consider the following 
$Z_2$ boundary conditions,
\begin{eqnarray}
& & \psi^{i_1,M_1}(-z)=\psi^{i_1,M_1}(z), \quad 
\psi^{i_2,M_2}(-z)=\psi^{i_2,M_2}(z),  \quad 
\psi^{i_3,M_3}(-z)=\psi^{i_3,M_3}(z) ,
\end{eqnarray}
for three types of zero-modes.
Then, we assign the first and second modes with left-handed 
and right-handed fermions, $L_i$ and $R_j$, 
while the third is assigned with Higgs fields $H_k$.
There are three $Z_2$ even modes for $M_1 = M_2 =4$, 
that is, the three generation model~\cite{Abe:2008fi,Abe:2008sx}, 
while there are five $Z_2$ even modes for $M_3=8$.
Their wavefunctions are shown in Table~\ref{tab:orbifold-model}.

\begin{table}[t]
\begin{center}
\begin{tabular}{|c|c|c|c|}\hline 
$i,j,k$ & $L_i $ & $R_j $ & 
$H_k $ 
\\ \hline \hline
0 & 
$\psi^{0,4}$ & $\psi^{0,4}$ & $\psi^{0,8}$ \\  
1 & 
$\frac{1}{\sqrt{2}}\left(\psi^{1,4}+\psi^{3,4}\right)$ & 
$\frac{1}{\sqrt{2}}\left(\psi^{1,4}+\psi^{3,4}\right)$ & 
$\frac{1}{\sqrt{2}}\left(\psi^{1,8}+\psi^{7,8}\right)$ \\  
2 & 
$\psi^{2,4}$ & 
$\psi^{2,4}$ & 
$\frac{1}{\sqrt{2}}\left(\psi^{2,8}+\psi^{6,8}\right)$ \\  
3 & - & - & 
$\frac{1}{\sqrt{2}}\left(\psi^{3,8}+\psi^{5,8}\right)$ \\  
4 & - & - & 
$\psi^{4,8}$ \\  \hline
\end{tabular}
\end{center}
\caption{Wavefunctions in the orbifold model.}
\label{tab:orbifold-model}
\end{table}

After orbifold projection, Yukawa couplings $Y_{ijk} L_i R_j H_k$ 
in this model are given by ~\cite{Abe:2008sx}
\begin{eqnarray}\label{eq:448}
Y_{ijk}H_k &=& 
\left( \begin{array}{ccc}
y_a H_0 + y_e H_4 & y_f H_3 + y_b H_1 & y_c H_2 \\
y_f H_3 + y_b H_1 & \frac{1}{\sqrt{2}}(y_a+y_e)H_2 
+ y_c(H_0+H_4) & y_b H_3+y_d H_1 \\
y_c H_2 & y_b H_3+y_d H_1 & y_e H_0 + y_a H_4 
\end{array} \right) .
\end{eqnarray}
Here, Yukawa coupling strengths, $y_a, y_b, \cdots, y_f$, 
are written as functions of moduli and they are, 
in general, different from each other.

We can take the basis of $L_i, R_j, H_k$ as 
eigenstates of $Z^2$ and $C^2$.
Such a basis is shown in Table~\ref{tab:z2c2}. 
Thus, if this effective theory has 
only $Z_2 \times Z_2 \times Z_2$ symmetry, 
the following couplings would be allowed,  
\begin{eqnarray}
Y_{ijk}H_k &=& 
\left( \begin{array}{ccc}
y_1 H_0+ y_2 H_2 +y_3 H_4   & 
y_4 H_1 + y_5 H_3 & 
y_6 H_0 + y_7 H_2 + y_8 H_4  \\
y_4' H_1 + y_5' H^3 &
y_9 (H_0+H_4) + y_{10} H_2 &
y_5' H_1 + y_4' H_3 \\ 
y_8 H_0 + y_7 H_2 +y_6 H_4 &
y_5  H_1 + y_4 H_3 &
y_3 H_0 + y_2 H_2 + y_1 H_4
\end{array} \right),  
\end{eqnarray}
where coupling strengths like $y_1$,$y_2$, etc. 
are independent parameters.
For example, the $Z_2 \times Z_2 \times Z_2$ 
symmetry allows non-vanishing couplings 
of $y_2$, $y_6$ and $y_8$.
However, these couplings are forbidden 
by the symmetry $Z_4 \ltimes (Z_4 \times Z_4) $ 
and such couplings do not appear in Eq.~(\ref{eq:448}).
Thus, Yukawa couplings derived from orbifolding 
are constrained more compared with 
the model, which has only the $Z_2 \times Z_2 \times Z_2$ 
flavor symmetry.

\begin{table}[t]
\begin{center}
\begin{tabular}{|c|cc|c|cc|c|cc|} \hline
$L_i$  & $Z^2$ & $C^2$ & $R_j $& $Z^2$ & $C^2$ 
& $H_k$ & $Z^2$ & $C^2$ \\ \hline \hline
$\frac{1}{\sqrt{2}} (L^0+L^2)$ &  1 &  1 &
$\frac{1}{\sqrt{2}} (R^0+R^2)$ &  1 &  1 &
$\frac{1}{\sqrt{2}} (H^0+H^4)$ &  1 &  1 \\
$\frac{1}{\sqrt{2}} (L^0-L^2)$ &  1 & --1 &
$\frac{1}{\sqrt{2}} (R^0-R^2)$ &  1 & --1 &
$\frac{1}{\sqrt{2}} (H^0-H^4)$ &  1 & --1 \\
$L_1$                          & --1 &  1 &
$R_1$                          & --1 &  1 &
$\frac{1}{\sqrt{2}} (H^1+H^3)$ & --1 &  1 \\
--                            &  -- &  -- & 
--                            &  -- &  -- & 
$\frac{1}{\sqrt{2}} (H^1-H^3)$ & --1 & --1 \\
--                            &  -- &  -- & 
--                            &  -- &  -- & 
$H_2$                          &  1 &  1 \\ \hline
\end{tabular}
\end{center}
\caption{Eigenstates of $Z^2$ and $C^2$}
\label{tab:z2c2} 
\end{table}

Similarly, other orbifold models have more constraints 
at least at tree level 
compared with unbroken symmetry as a footprint 
of larger flavor symmetries before orbifolding.
Such a structure would be useful for 
phenomenological applications.

\section{Conclusion and discussion}

We have studied the non-abelian flavor symmetries, 
which can appear in magnetized brane models.
We have found that $D_4$, $\Delta(27)$ and other 
$Z_g \ltimes (Z_g \times Z_g)$ 
flavor symmetries can appear from magnetized brane models 
with non-vanishing Wilson lines.
Matter fields with several representations of 
these discrete flavor symmetries can appear.
When we consider vanishing Wilson lines, 
these flavor symmetries are enhanced like 
$D_4 \times Z_2$, $\Delta(54)$, etc.
These results are interesting to apply for model building 
of realistic quark/lepton mass matrices.
We have also discussed the flavor symmetry breaking 
on the orbifold background.

Since intersecting D-brane models are T-duals of 
magnetized brane models, we would obtain 
the same results in intersecting D-brane models.

It is important to study anomalies of non-abelian 
flavor symmetries.
If string theory leads to anomaly-free effective low-energy 
theories including discrete symmetries, 
anomalies of discrete symmetries must be canceled by 
the Green-Schwarz mechanism.
Those discrete anomalies were studied 
within the framework of heterotic orbifold models in 
\cite{Araki:2008ek}, and it was shown that 
discrete anomalies can be canceled by 
the Green-Schwarz mechanism.
Furthermore, important relations of discrete anomalies 
with U(1) anomalies and others were found.
(See also \cite{Araki:2007ss}.)
It is important to extend such an analysis to 
magnetized/intersecting brane models.

\subsection*{Acknowledgement}

K.-S.~C., T.~K. and H.~O. are supported in part by the Grant-in-Aid for 
Scientific Research No.~20$\cdot$08326, No.~20540266 and
No.~21$\cdot$897 from the 
Ministry of Education, Culture, Sports, Science and Technology of Japan.
T.~K. is also supported in part by the Grant-in-Aid for the Global COE 
Program "The Next Generation of Physics, Spun from Universality and 
Emergence" from the Ministry of Education, Culture,Sports, Science and 
Technology of Japan.

\end{document}